\documentclass[aps,prb,amsmath,amssymb,nobibnotes,epsfig,jmp]{revtex4-1}

\usepackage{bm}
\usepackage{latexsym}
\usepackage{amsthm}
\usepackage{graphicx}
\usepackage{subeqnarray}

\newtheoremstyle{jmp}
{\topsep}{\topsep}
{\normalfont\itshape}
{\parindent}{\normalfont\bfseries}
{:}{.5em}
{}

\theoremstyle{jmp}
\newtheorem{thm}{Theorem}

\makeatletter
\renewcommand\frontmatter@title@above{}
\renewcommand\frontmatter@title@format{\large\bfseries\raggedright}
\renewcommand\frontmatter@title@below{\addvspace{6\p@}}
\makeatother

\makeatletter
\renewcommand\frontmatter@authorformat
{\preprintsty@sw{\vskip0.5pc\relax}{}
\@tempskipa\@flushglue
\@flushglue\z@ plus50\p@\relax
\raggedright\advance\leftskip20mm\relax
\@flushglue\@tempskipa
\parskip\z@skip}
\makeatother

\makeatletter
\renewcommand\frontmatter@affiliationfont
{\small\slshape\selectfont\baselineskip10.5\p@\relax
\@tempskipa\@flushglue
\@flushglue\z@ plus50\p@\relax
\raggedright\advance\leftskip20mm\relax
\@flushglue\@tempskipa}
\makeatother

\makeatletter
\renewcommand{\section}{\@startsection{section}{1}{\z@}
{0.8cm \@plus1ex minus .2ex}{0.5cm}{\normalfont\small\bfseries}}
\renewcommand{\subsection}{\@startsection{subsection}{2}{-12pt}
{0.8cm \@plus1ex minus .2ex}{0.5cm}{\normalfont\small\bfseries}}
\renewcommand{\subsubsection}{\@startsection{subsubsection}{3}{-12pt}
{0.8cm \@plus1ex minus .2ex}{0.5cm}{\normalfont\small\itshape}}
\makeatother

\makeatletter
\renewcommand\bibsection
{\section*{}
\@nobreaktrue}
\makeatother

\newcommand{\nc}{\newcommand*}

\nc{\reff}[1]{(\ref{#1})}
\nc{\nn}{\nonumber}
\nc{\ts}{\textstyle}
\nc{\ds}{\displaystyle}

\def\basn{\left\{\Psi_{n}(x)\right\}_{n=0}^{\infty}}

\def\sAn{\left\{A_{n}^{(j))}\right\}_{n=0}^{\infty}}

\begin{document}

\title{Comment on "On the dimensions of the oscillator algebras induced by orthogonal polynomials"
[J. Math. Phys. {\bf 55}, 093511 (2014)]}
\thanks{This work was done under the partial support of the RFBR grant 15-01-03148-à}

\author{V. V. Borzov}
\email{borzov.vadim@yandex.ru}
\affiliation{Department of Mathematics, St.Petersburg State University of Telecommunications, 191065, Moika  61, St.Petersburg, Russia}

\author{E. V. Damaskinsky}
\email{evd@pdmi.ras.ru}
\affiliation{Department of Natural Sciences, Institute of Defense Technical Engineering (VITI),\\ 191123, Zacharievskaya 22, St.Petersburg, Russia}

\date{01.03.2015}

\begin{abstract}
\noindent
In the interesting paper G. Honnouvo and K. Thirulogasanthar  [J. Math. Phys. {\bf 55} , 093511 (2014)]
the authors obtained the necessary and sufficient conditions under which  the oscillator algebra connected
with orthogonal polynomials on real line is finite-dimensional (and in this case the dimension of the algebra is
always equal four). In the cited article, only the case when polynomials are orthogonal with respect to a
symmetric measure on the real axis was considered.
Unfortunately, the sufficient condition from this paper  is incomplete.
Here we clarify the sufficient part of the corresponding theorem from that paper
and extend the results to the case when measure is not symmetric.
\end{abstract}

\maketitle

\setlength{\parindent}{4mm}

{\bf Introduction.}
The basic object of the paper  \cite{01} is the notion of generalized oscillator
connected with the family of orthogonal polynomials introduced in the article \cite{02}.
We briefly recall this definition.
Let $\mu$ be a probability measure on ${\mathbb R}$ with finite moments
$\mu_n=\int_{-\infty}^{\infty} x^n\,d\mu$.
These moments uniquely define the real sequences
$\left\{b_{n}\right\}_{n=0}^{\infty}$,$\left\{a_{n}\right\}_{n=0}^{\infty}$
and the system of orthogonal polynomials with recurrent relations ($n\geq0$)
\begin{equation}\label{1}
x\,P_{n}(x)=b_{n}\,P_{n+1}(x)+a_{n}P_{n}(x)+b_{n-1}P_{n-1}(x);\quad P_{0}(x)=1,\quad b_{-1}=0.
\end{equation}
These polynomials form an orthonormal basis in the Hilbert space
$\mathcal{H}_{\mu}=\text{L}^2\left(\mathbb{R};\mu\right)$. We must distinguish two cases:

1) $a_n=0$ --- the symmetric case;

2) $a_n\neq0$ --- the nonsymmetric case.

We define the ladder operators $a^{\pm}$ and the number operator $N$ in $\mathcal{H}$
by formulas
\begin{align}
a^+ P_{n}(x)&=\sqrt{2}b_n P_{n+1}(x);\nn \\
a^- P_{n}(x)&=\sqrt{2}b_{n-1}P_{n-1}(x),\quad n\geq0; \\
N P_{n}(x)&=n P_{n}(x).\nn
\end{align}
Next, we denote by $B(N)$ a function of selfadjoint operator $N$ in the space $\mathcal{H}$.
This function  acts on the basis vectors $\basn$ as follows
\begin{equation}
B(N) P_{n}(x)=b_{n-1}^{\,\, 2} P_{n}(x),\qquad
B(N+I) P_{n}(x)=b_{n}^{\, 2} P_{n}(x),\quad n\geq0.
\end{equation}
The following result was proven in \cite{02}.
\begin{thm}\label{2}
The operators $a^{\pm},\, N,\, I$ obey the following relations
\begin{equation}\label{3}
a^-a^+=2B(N+I), \quad a^+a^-=2B(N),\quad [N,a^{\pm}]=\pm a^{\pm}.
\end{equation}
\end{thm}

{\bf Definition}.
{\it An associative algebra $\mathfrak{A}$ is called the generalized oscillator algebra
corresponding to the orthogonal polynomial system $\basn$, which satisfy recurrent relations \reff{1},
if $\mathfrak{A}$ is generated by
the operators $a^{\pm},\, N,\, I$, that obey the relations \reff{3}}.

{\bf Remark 1}. Let $\mathfrak{A_{a}}$ be the generalized oscillator algebra corresponding
to the recurrent relations \reff{1} in nonsymmetric case  ($a_n\neq0$), and
$\mathfrak{A_{s}}$ - the generalized oscillator algebra for related symmetric case.
 As follows from above definition the two algebras coincide and therefore their dimensions equal,
 i.e.  $\dim {\mathfrak{A_{a}}}= \dim {\mathfrak{A_{s}}}$.

{\bf The dimension of generalized oscillator algebras}.
In a recently published paper \cite{01} the authors investigated the conditions
wherein algebra of the generalized oscillator $\mathfrak{A}$, associated with
orthogonal polynomials in the manner described above,
is finite-dimensional. In \cite{01} it was considered {\bf only} the case of orthogonal polynomials
for a symmetric measure on the real axis (when the Jacobi matrix corresponding to the recurrent
relations \reff{1} has zero diagonal).
The following results were proven in \cite{01}.

\begin{thm}\label{4}
 Let us define the sequence
\begin{equation}\label{5}
A_n^{(0)}=b_{n}^{\, 2}-b_{n-1}^{\, 2},\ldots,A_n^{(j)}=A_{n+1}^{(j-1)}-A_n^{(j-1)},
\end{equation}
$j=1,2,\ldots,\quad n=0,1,\ldots$.
If for any fixed $j>0$, the sequence $\sAn$ is not constant, i.e.
$A_n^{(j)}\neq$const, $n=0,1,\ldots$, then the generalized oscillator algebra $\mathfrak{A}$
is infinite dimensional.
\end{thm}

\begin{thm}\label{6}
The generalized oscillator algebra $\mathfrak{A}$
is of finite dimension if and only if
\begin{equation}\label{7}
 b_{n}^{\, 2}=a_0+a_1n+a_2n^2,\qquad a_0, a_1, a_2\in\mathbb{R}.
\end{equation}
In this case the dimension of the algebra $\mathfrak{A}$ is four.
\end{thm}

{\bf Remark 2 }. Unfortunately, the only a necessary part of the above theorem from
the paper \cite{01} is correct.
In order to make the right also a sufficient condition of theorem
we clarified in \cite{03} the condition \reff{7}.
Namely, the coefficients in \reff{7} must satisfy the following equality
\begin{equation}\label{8}
    a_1=a_0+a_2.
\end{equation}
\newpage
{\bf The proof of correct sufficient condition in the
theorem 3}.
\medskip

Let the coefficients  $b_{n}^{\, 2}$ have the following form,
\begin{equation}\label{9}
 b_{n}^{\, 2}=\sum_{i=0}^{p}a_in^i,\qquad a_0, a_1, a_2\in\mathbb{R}\qquad p=0,1,2,
\end{equation}
and let $\mathfrak{A}(p)$ denotes the corresponding generalized oscillator algebra.
 We proof the sufficient condition by cases.
\medskip

{\bf The case p=0}:
\medskip

If $b_{n}^{\, 2}=constant\neq{0}$, n=0,1,2,..., (but $ b_{-1}=0$ ) then we have
\begin{equation}\label{10}
[a^-,a^+]\,P_n=2\,\delta(n)\,b_{0}^{\, 2}\, P_n, \quad [N,a^{\pm}]=\pm a^{\pm}.
\end{equation}
From \reff{5} and \reff{10} we have
$A_n^{(j)}=(-1)^j\,\delta(n)\,b_{0}^{\, 2}\neq constant,\quad n=0,1,\ldots$,
for every fixed $j>0$.
Thus, the algebra $\mathfrak{A}(0)$ is of infinite dimension.
\medskip

{\bf The case p=1}:
\medskip

If $b_{n}^{\, 2}=n+a_0$, n=0,1,2,..., (but $ b_{-1}=0$ ) then we have
\begin{equation}\label{11}
[a^-,a^+] P_n=2(1+\delta(n)(a_0-1)) P_n, \quad [N,a^{\pm}]=\pm a^{\pm}.
\end{equation}
Thus, the dimension of algebra $\mathfrak{A}(1)$ is finite (namely, it equals to four) if and only if $a_0=1$.
\medskip

{\bf The case p=2}:
\medskip

If $b_{n}^{\, 2}=n^2+a_1 n+a_0$, n=0,1,2,..., (but $ b_{-1}=0$ ) then we have
\begin{equation}\label{12}
[a^-,a^+]P_n=2n-1+a_1+\delta(n)\,(-2n+1-a_1+a_0))\,P_n, \quad [N,a^{\pm}]=\pm a^{\pm}.
\end{equation}
Thus, the dimension of algebra $\mathfrak{A}(2)$ is finite (namely, it equals to four)
if and only if $a_1=a_0+1$.

It is clear that in general case, when  $b_{n}^{\, 2}$ is defined by \reff{7},
the dimension of corresponding algebra $\mathfrak{A}$ is finite (namely, it equals to four)
if and only if the condition \reff{8} is true.

Finally, in the \cite{01} it was proved that the algebra $\mathfrak{A}(p)$ is
of infinite dimension for $p\geq3$.

As a result, we have obtained the correct Theorem 3.
\begin{thm}\label{13}
The generalized oscillator algebra $\mathfrak{A}$  is of finite dimension if  and only if
\begin{equation}\label{14}
 b_{n}^{\, 2}=(a_0+a_2n)(1+n),\qquad a_0, a_2\in\mathbb{R}.
\end{equation}
In this case the dimension of the algebra $\mathfrak{A}$ is four.
\end{thm}

{\bf Remark 3}.  It follows from  Remark 1 that Theorem 2 and Theorem 4
remain true for arbitrary orthogonal polynomials on real axis.

As an example illustrating the last theorem, we consider the Laguerre polynomials satisfying nonsymmetric recurrent relations.
These polynomials
\begin{equation*}
L_n^{\alpha}(x)=\frac{\alpha+1}{n!}{}_1F_1(-n,\alpha+1;x).
\end{equation*}
are orthogonal in the Hilbert space
$\mathcal{H}=\text{L}^2(\mathbb{R}_+^1;x^{\alpha}\exp(-x)\text{d}x)$.
Normalized polynomials
\begin{equation*}
\Psi_n(x)\!=\!d_n^{-1}L_n^{\alpha}(x),\quad d_n\!=
\!\sqrt{\frac{\Gamma(n\!+\!\alpha\!+\!1)}{n!}},\,\, n\geq0
\end{equation*}
fulfill the nonsymmetric recurrence relations  \reff{2} with
\begin{equation*}
b_n\!=\!-\sqrt{(n\!+\!1)(n\!+\!\alpha\!+\!1)},\quad a_n\!=\!2n\!+\!\alpha\!+\!1.
\end{equation*}
In this case $b_n^{\,\, 2}$ have the form \reff{14}, and hence the corresponding algebra
$\mathfrak{A}_L$  is a four-dimensional but not isomorphic
to the algebra of the harmonic oscillator.

{\bf In conclusion}, we note that in the work \cite{03} we have proved that the
oscillator algebra associated with a system of
 orthogonal polynomials on the real line is finite-dimensional (namely, four-dimensional)
 {\bf only} for the case of Hermite or Laguerre polynomials or for those polynomials,
 which generate the oscillator algebra that is isomorphic to the above mentioned algebras.
\bigskip

{\bf Acknowledgements}. The research of EVD performed with the financial support of RFBR,
grant 15-01-03148-a.

\end{document}